\documentclass[prb,twocolumn,superscriptaddress,preprintnumbers]{revtex4-2}

\usepackage{graphicx}  
\usepackage{subfigure}
\usepackage{multirow}
\usepackage{ulem}
\usepackage{fancyhdr}
\usepackage{longtable}

\usepackage[T1]{fontenc}
\usepackage{dcolumn}   

\usepackage{bm}        
\usepackage{amsfonts}  
\usepackage{amsmath}   
\usepackage{amssymb}   
\usepackage{makecell} 
\usepackage[colorlinks]{hyperref}

\hypersetup{
citecolor={blue},
urlcolor={blue}
}
\usepackage{array} 
\usepackage{ragged2e} 

\begin{document}

\title{Resolving phonon-mediated superconducting pairing symmetries from first-principles calculation}

\author{Zimeng Zeng}
\affiliation{State Key Laboratory of Low Dimensional Quantum
Physics, Department of Physics, Tsinghua University, Beijing
100084, China}

\author{Xiaoming Zhang}
\affiliation{College of Physics and Optoelectronic Engineering, Ocean University of China, Qingdao, Shandong 266100, China}

\author{Shunhong Zhang}
\affiliation{The International Center for Quantum Design of Functional Materials (ICQD),
 University of Science and Technology of China, Hefei 230026, China}
 \affiliation{Hefei National Laboratory, University of Science and Technology of China, Hefei, Anhui 230088, China}

\author{Jian Wu}
\email{wu@tsinghua.edu.cn}
\affiliation{State Key Laboratory of Low Dimensional Quantum
Physics, Department of Physics, Tsinghua University, Beijing
100084, China}
\affiliation{Frontier Science Center for Quantum Information, Beijing, China.}

\author{Zheng Liu}
\email{zliu23@buaa.edu.cn}
\affiliation{School of Physics, Beihang University, Beijing
100191, China}
\date{\today}

\begin{abstract}  
The quest for topological superconductors triggers revived interests in resolving non-$s$-wave pairing channels mediated by phonons. While density functional theory and density functional perturbation theory have established a powerful framework to calculate electron-phonon couplings in real materials in a first-principles way, its application is largely limited to conventional $s$-wave superconductivity. Here, we formulate an efficient and simple-to-use algorithm for first-principles pairing channel analysis, and apply it to several representative material systems.

\end{abstract}

\maketitle 
\section{Introduction}

In the past two decades, the first-principles calculation within the framework of density functional theory (DFT) has reached a status to reliably describe not only the normal states of a wide range of materials, but also conventional superconductivity (SC) mediated by phonons~\cite{RevModPhys.89.015003}. Calculating the electron-phonon couplings (EPCs) from first principle is rapidly reaching maturity, thanks to the development of density functional perturbation theory (DFPT)~\cite{RevModPhys.73.515}.

It is commonly known that the predominating pairing channel from phonons is of a $s$-wave symmetry, and the Cooper pairs form spin singlets. In practice, the spin degrees of freedom of EPCs are thus ignored in most first-principles calculations. To predict the superconducting transition temperature (T$_c$), the $s$-wave channel is implicitly assumed, either in McMillan-Allen-Dynes formula~\cite{mcmillan1968transition} or the self-consistent solution of Migdal-Eliashberg equations~\cite{eliashberg1960interactions}.

Recently, the potential role of phonons in achieving unconventional superconductivity attracts revived theoretical attentions. The concern is that while electron-electron Coulomb repulsion always suppresses the conventional $s$-wave pairing, it can be collaborative with phonons in stabilizing other pairing symmetries, e.g., in a topological superconductor~\cite{fu2010odd,Wan,brydon2014odd}, or in high-$T_c$ cuprate and pnictides superconductors~\cite{PhysRevB.104.L140506,PhysRevB.83.092505,shen2002role}. In these cases, the non-$s$-wave pairing channels from EPC should be taken into account.

First-principles analysis along this direction is very limited. The pioneering calculation by Wan and Savrasov~\cite{Wan} on doped Bi$_2$Se$_3$ employed a set of orthogonal polynomials defined on the Fermi surface (FS), known as FS harmonics~\cite{PhysRevB.13.1416}, to project EPC to different pairing channels. The results clearly indicate that non-$s$-wave pairing information is readily encoded in the standard DFPT EPC data. However, the construction of FS harmonics is tedious, especially for complicated FS's. The functions depend on not only the crystal structure, but also FS geometry, which hinders high-throughput applications of the method.

Nomoto $et$ $al.$~\cite{Tokura} extended DFT for superconductivity (SCDFT)~\cite{PhysRevB.72.024545,PhysRevB.72.024546} to deal with odd-parity pairing in an \textit{ab initio} way. The treatment is exact when $S_z$ is a good quantum number, e.g., without the spin-orbit coupling (SOC). When SOC is strong, however, the exact forms of odd-parity pairing in the basis of Kohn-Sham (KS) eigenstates are not known \textit{a priori}.

The primary goal of this article is to formulate an efficient and simple-to-use algorithm for first-principles EPC pairing-channel analysis. The key is to properly solve gauge  indeterminacy of the EPCs. After that, we show that the pairing channels can be resolved by standard matrix diagonalization. We call this approach the ``direct diagonalization method'' (DDM). In Sec. II, we will first review the general theory of phonon-mediated pairing interactions, and then discuss the gauge complexity. In the end, we present the DDM. In Sec. III, the DDM is applied to several realistic material systems. Section IV concludes this article. 

\section{Theoretical framework}

\subsection{General theory}

We start by considering a system with inversion ($\mathcal{I}$) and time-reversal ($\mathcal{T}$) symmetries. Without $\mathcal{I}$, the definition of parity is missing, and the gap function is in general a mixture of odd and even parities. Without $\mathcal{T}$, the scenario is complicated by the coexistence of magnetism, which we defer for future works. The combination of $\mathcal{I}$ and $\mathcal{T}$ enforces that every eigenstate is at least doubly degenerate. Every KS eigenstate can then be labeled by a band index $n$, the lattice momentum $\mathbf{k}$ and a pseudospin variable $s=\pm$.

The DFT+DFPT calculations quantify the following EPC Hamiltonian:
\begin{eqnarray}
H_{ep}=\sum_{n,m,\mathbf{k,q},s,s'\upsilon}g_{n,\mathbf{k},s;m,\mathbf{k+q},s'}^{\upsilon}c_{m,\mathbf{k+q},s'}^{\dagger}c_{n,\mathbf{k},s}(b_{\upsilon,-\mathbf{q}}^{\dagger}+b_{\upsilon ,\mathbf{q}}), \nonumber
\end{eqnarray}
in which $c_{n,\mathbf{k},s}$ ($b_{\upsilon,\mathbf{q}}$) is the annihilation operator of the electron (phonon)  eigenstate. The phonons are labeled by a mode index $\upsilon$ and the lattice vector $\mathbf{q}$. $g_{n,\mathbf{k},s;m,\mathbf{k+q},s'}^{\upsilon}$ is the EPC coefficient.

Within the second-order perturbation, the EPC generates the pairing interaction:
\begin{eqnarray}\label{eq:Hpair}
H_{pair}=-\frac{1}{2}\sum_{n,\mathbf{k},m,\mathbf{k}',s_{1,2,3,4}}V_{s_1,s_2;s_3,s_4}(n,\mathbf{k};m,\mathbf{k}') \\ \nonumber
c_{m,\mathbf{k}',s_3}^{\dagger}c_{m,\mathbf{-k}',s_4}^{\dagger}c_{n,\mathbf{-k},s_2}c_{n,\mathbf{k},s_1},
\end{eqnarray}
with
\begin{eqnarray}\label{eq:V}
    V_{s_1,s_2,s_3,s_4}(n,\mathbf{k};m,\mathbf{k}')&=&2\sum_\upsilon \frac{g_{n,\mathbf{k},s_1;m,\mathbf{k'},s_3}^{\upsilon}g_{n,\mathbf{-k},s_2;m,\mathbf{-k'},s_4}^{\upsilon}}{\omega_{\mathbf{k'}-\mathbf{k},\upsilon}}  \nonumber \\
    &\times&\delta(\epsilon_{n,\mathbf{k}}-\epsilon_F)\delta(\epsilon_{m,\mathbf{k}’}-\epsilon_F),
\end{eqnarray}
in which $\epsilon_{n,\mathbf{k}}$ ($\omega_{\mathbf{k'}-\mathbf{k},\upsilon}$) is the electron (phonon) eigen-energy, and we restrict the involved electrons residing at the Fermi energy ($\epsilon_F$). We do not consider strong coupling systems beyond this treatment.

To bridge the interaction to a matrix analysis problem, it is convenient to view $V_{s_1,s_2;s_3,s_4}(n,\mathbf{k};m,\mathbf{k}')$ as the components of a matrix $\mathbb{V}$ and define a vector operator $\hat \Delta$ with its components 
\begin{eqnarray}\label{eq:Delta}
   \hat \Delta(n,\mathbf{k},s_1,s_2)=c_{n,\mathbf{-k},s_2}c_{n,\mathbf{k},s_1}, 
\end{eqnarray}
which physically annihilates a Copper pair consisting of electrons at the $|n,\mathbf{k},s_1\rangle$ and $|n,\mathbf{-k},s_2\rangle$ states. Its ground-state expectation value $\langle \hat \Delta(n,\mathbf{k},s_1,s_2)\rangle$ reflects the SC order parameter. Eq.~(\ref{eq:Hpair}) is then transformed into a compact quadratic form:
\begin{eqnarray}
    H_{pair}=-\frac{1}{2}\hat \Delta^\dagger\mathbb{V}\hat \Delta.
\end{eqnarray}
Roughly, diagonalizing the kernel $\mathbb{V}$ is supposed to decompose different pairing channels. 

\subsection{Gauge complexity}

The complexity is that EPCs are routinely evaluated on a $k$-mesh in the full Brillouin zone, where KS orbitals at each $k$ point are treated independently with ''gauge'' indeterminacy. It is straightforward to see that $\hat \Delta(n,\mathbf{k},s_1,s_2)$ is gauge dependent by adding an arbitrary U(1) phase to $c_{n,\mathbf{k},s_1}$. More critically, the double degeneracy allows any recombination of the degenerate KS orbitals as the eigenstate. Therefore, an arbitrary SU(2) rotation is also presented in the first-principles data, especially for strong SOC systems, which hinders a textbook definition of the pairing symmetry assuming $s_1,s_2$ in $\hat \Delta$ refer to a global spin axis. It is worth mentioning that maximally localized Wannier function interpolation is a powerful procedure for \textit{ab initio} gauge fixing~\cite{RevModPhys.84.1419}, but to properly specify the pseudospin index, especially for strong SOC systems, could still be tricky.

\subsection{DDM solution}

The main advantage of the DDM is to avoid specifying the gauge condition, such that the computational procedure could label the pairing channel in any gauge. Let us denote
\begin{eqnarray}\label{eq:T}
\mathcal{T}c_{n,\mathbf{k},s}\equiv s\tilde{c}_{n,-\mathbf{k},-s},    
\end{eqnarray}
in which the sign $s$ before $\tilde{c}$ is used to explicitly track the $\mathcal{T}^2=-1$ condition for an electron. The tilde symbol implies that $(\tilde{c}_{n,-\mathbf{k},+}, \tilde{c}_{n,-\mathbf{k},-})$ is allowed to differ from $(c_{n,-\mathbf{k},+}, c_{n,-\mathbf{k},-})$, i.e., the numerical KS states at $-\mathbf{k}$ without gauge fixing, by an arbitrary U(1)$\times$SU(2) transformation. Thus defined $\tilde{c}_{n,-\mathbf{k},-s}$ also relates to $c_{n,\mathbf{k},-s}$ via spatial inversion as:
\begin{eqnarray}   \label{eq:I} \mathcal{I}c_{n,\mathbf{k},-s}=e^{i\alpha_{n,\mathbf{k}}}\tilde{c}_{n,-\mathbf{k},-s}.
\end{eqnarray}
Note that theoretical analysis usually implicitly assumes $e^{i\alpha_{n,\mathbf{k}}}=1$, which however does not automatically hold for the numerical KS states. A U(1) transformation $c_{n,\mathbf{k},\pm}\rightarrow e^{i\alpha_{n,\mathbf{k},\pm}}c_{n,\mathbf{k},\pm}$ will lead to an overall phase shift $\alpha_{n,\mathbf{k}}=\alpha_{n,\mathbf{k},+}+\alpha_{n,\mathbf{k},-}$ in Eq. (6).

By switching to the $\tilde{c}$ operators for all the -$\mathbf{k}$ states, we rewrite Eqs. (\ref{eq:Hpair}) and (\ref{eq:V}) into
\begin{eqnarray}
H_{pair}=-\frac{1}{2}\sum_{n,\mathbf{k},m,\mathbf{k}',s_{1,2,3,4}}V_{s_1,s_2;s_3,s_4}(n,\mathbf{k};m,\mathbf{k}') \\ \nonumber
c_{m,\mathbf{k}',s_3}^{\dagger}\tilde{c}_{m,\mathbf{-k}',s_4}^{\dagger}\tilde{c}_{n,\mathbf{-k},s_2}c_{n,\mathbf{k},s_1},   
\end{eqnarray}
and
\begin{eqnarray}
   V_{s_1,s_2;s_3,s_4}(n,\mathbf{k};m,\mathbf{k}')=& \nonumber\\ 2\sum_\upsilon s_2&s_4\frac{g_{n,\mathbf{k},s_1;m,\mathbf{k'},s_3}^{\upsilon}(g_{n,\mathbf{k},-s_2;m,\mathbf{k'},-s_4}^{\upsilon})^*}{\omega_{\mathbf{k'}-\mathbf{k},\upsilon}}  \nonumber \\
    &\times \delta(\epsilon_{n,\mathbf{k}}-\epsilon_F)\delta(\epsilon_{m,\mathbf{k}’}-\epsilon_F).
\end{eqnarray}
The emergence of summand sign $s_2,s_4$ is a direct consequence of $\mathcal{T}^2=-1$ [cf. Eq. (\ref{eq:T})]. A useful point is that the EPC between $\tilde c_{m,\mathbf{-k}',s'}^{\dagger}$ and $\tilde c_{n,\mathbf{-k},s}$ can be readily obtained from that between $c_{m,\mathbf{k}',-s'}^{\dagger}$ and $ c_{n,\mathbf{k},-s}$ by applying $\mathcal{T}$, so the tilde states never need to be explicitly constructed.

Accordingly, Eq. (\ref{eq:Delta}) is rewritten as 
\begin{eqnarray}\label{eq:newDelta}
\hat\Delta(n,\mathbf{k},s_1,s_2)=\tilde{c}_{n,\mathbf{-k},s_2}c_{n,\mathbf{k},s_1}.
\end{eqnarray}
This definition greatly facilitates numerical analysis of gap function symmetry, because its transformation under $\mathcal{I}$ is well-defined. Referring to Eq. (\ref{eq:I}), we have
\begin{eqnarray}  
\mathcal{I}\hat\Delta(n,\mathbf{k},s_2,s_1)=-\hat\Delta(n,\mathbf{k},s_1,s_2). \\ \nonumber
\end{eqnarray}
Note that the undetermined U(1) phase $e^{i\alpha_{n,\mathbf{k}}}$ cancels.

Now, we can construct one even-parity operator:
\begin{eqnarray}
\hat\Delta^S(n,\mathbf{k})=\frac{1}{\sqrt{2}}[\hat\Delta(n,\mathbf{k},+,-)-\hat\Delta(n,\mathbf{k},-,+)],
\end{eqnarray}
and three odd-parity operators
\begin{eqnarray}
\hat\Delta^{T_0}(n,\mathbf{k})&=&\frac{1}{\sqrt{2}}[\hat\Delta(n,\mathbf{k},+,-)+\hat\Delta(n,\mathbf{k},-,+)] \nonumber \\
\hat\Delta^{T_{\pm}}(n,\mathbf{k})&=&\hat\Delta(n,\mathbf{k},\pm,\pm).
\end{eqnarray}
Transforming into this basis, the kernel $\mathbb{V}$ is block diagonalized into the even-parity singlet sector ($\mathbb{V}^S$) and the odd-parity triplet sector ($\mathbb{V}^T$). The matrix elements are:
\begin{eqnarray}
    V^S(n,\mathbf{k};m,\mathbf{k}')&=&\sum_{\upsilon ,s_1,s_3}\frac{|g_{n,\mathbf{k},s_1;m,\mathbf{k'},s_3}^{\upsilon}|^2}{\omega_{\mathbf{k'}-\mathbf{k},\upsilon}}  \nonumber \\
    &\times&\delta(\epsilon_{n,\mathbf{k}}-\epsilon_F)\delta(\epsilon_{m,\mathbf{k}’}-\epsilon_F), 
\end{eqnarray}
and
\begin{widetext}
\begin{eqnarray}\label{eq:VT}
    V^T(n,\mathbf{k};m,\mathbf{k}')&=& 2\sum_\upsilon 
    \begin{pmatrix}
    g_{n,\mathbf{k},+;m,\mathbf{k'},+}^{\upsilon}(g_{n,\mathbf{k},-;m,\mathbf{k'},-}^{\upsilon})^* & -\sum_s \frac{sg_{n,\mathbf{k},+;m,\mathbf{k'},s}^{\upsilon}(g_{n,\mathbf{k},-;m,\mathbf{k'},s}^{\upsilon})^*}{\sqrt{2}} & -g_{n,\mathbf{k},+;m,\mathbf{k'},-}^{\upsilon}(g_{n,\mathbf{k},-;m,\mathbf{k'},+}^{\upsilon})^* \\
    -\sum_s \frac{sg_{n,\mathbf{k},s;m,\mathbf{k'},+}^{\upsilon}(g_{n,\mathbf{k},s;m,\mathbf{k'},-}^{\upsilon})^*}{\sqrt{2}} & \sum_{ ss'}\frac{ss'|g_{n,\mathbf{k},s;m,\mathbf{k'},s'}^{\upsilon}|^2}{2} & \sum_s \frac{sg_{n,\mathbf{k},s;m,\mathbf{k'},-}^{\upsilon}(g_{n,\mathbf{k},s;m,\mathbf{k'},+}^{\upsilon})^*}{\sqrt{2}} \\
    -g_{n,\mathbf{k},-;m,\mathbf{k'},+}^{\upsilon}(g_{n,\mathbf{k},+;m,\mathbf{k'},-}^{\upsilon})^* & \sum_s \frac{sg_{n,\mathbf{k},-;m,\mathbf{k'},s}^{\upsilon}(g_{n,\mathbf{k},+;m,\mathbf{k'},s}^{\upsilon})^*}{\sqrt{2}} & g_{n,\mathbf{k},-;m,\mathbf{k'},-}^{\upsilon}(g_{n,\mathbf{k},+;m,\mathbf{k'},+}^{\upsilon})^*
    \end{pmatrix} \nonumber \\ 
    &\times&\frac{\delta(\epsilon_{n,\mathbf{k}}-\epsilon_F)\delta(\epsilon_{m,\mathbf{k}’}-\epsilon_F)}{\omega_{\mathbf{k'}-\mathbf{k},\upsilon}}.
\end{eqnarray}
\end{widetext}

It is heuristic to inspect that the average of $V^S(n,\mathbf{k};m,\mathbf{k}')$ normalized by the Fermi surface density of states is the isotropic dimensionless EPC strength $\lambda$ commonly used to predict the $s$-wave $T_c$~\cite{RevModPhys.89.015003}, which implicitly assumes that the matrix components do not contain any singular variation. The weak-SOC approximation employed by Nomoto $et$ $al.$ \cite{Tokura} to evaluate odd-parity pairing can be viewed as neglecting the off-diagonal matrix elements in Eq. (\ref{eq:VT}).

Without introducing these approximations, we proceed by directly diagonalizing $\mathbb{V}^{S(T)}$:
\begin{eqnarray}
    \mathbb{V}^{S}|\sigma_i\rangle&=&\lambda^S_i|\sigma_i\rangle, \nonumber \\
    \mathbb{V}^{T}|\tau_i\rangle&=&\lambda^T_i|\tau_i\rangle.   
\end{eqnarray}
In practice, we non-dimensionalize $\mathbb{V}$ by $V_{s_1,s_2;s_3,s_4}(n,\mathbf{k};m,\mathbf{k}')\rightarrow V_{s_1,s_2;s_3,s_4}(n,\mathbf{k};m,\mathbf{k}')N_{n,\mathbf{k}}$, where $N_{n,\mathbf{k}}$ is the contribution of the $|n,\mathbf{k},s\rangle$ state to the total density of states. This makes the largest $\lambda^S_i$, which we denote as $\lambda^S_{max}$, directly comparable to the isotropic dimensionless EPC strength $\lambda$ reported in literature. The largest $\lambda^T_i$ ($\lambda^T_{max}$) gives the leading odd-parity spin-triplet pairing strength.

The eigenvectors:
\begin{eqnarray}\label{eq:eigenvector}
|\sigma_i\rangle&=&[\xi_i(n_1,\mathbf{k}_1), \xi_i(n_1,\mathbf{k}_2), ...]^T, \\
|\tau_i\rangle&=&[\xi_i(T_0,n_1,\mathbf{k}_1), \xi_i(T_+,n_1,\mathbf{k}_1), \xi_i(T_-,n_1,\mathbf{k}_1),  ...]^T \nonumber
\end{eqnarray}
reflect the $\mathbf{k}$-space symmetry of the corresponding pairing channel, which can be classified by the irreducible representations (irreps) of the symmetry group.

There is still one remaining issue. So far, $\hat\Delta(n,\mathbf{k},s_1,s_2)=\tilde{c}_{n,\mathbf{-k},s_2}c_{n,\mathbf{k},s_1}$ and $\hat\Delta(n,-\mathbf{k},s_2,s_1)=\tilde{c}_{n,\mathbf{k},s_1}c_{n,\mathbf{-k},s_2}$ have been treated as independent operators, which are actually related by an undecided gauge transformation and the fermion anti-commutation relation. (Mathematically, the basis of $\mathbb{V}$ is overcomplete.) Therefore, additional constraints should be applied to the eigenvectors. 

Specifically, the symmetrized even-parity operator $\hat\Delta^S(n,\mathbf{k})$ is U(1) and SU(2) invariant. The constraint to $|\sigma_i\rangle$ is simply $\xi_i(n_m,\mathbf{k}_j)=\xi_i(n_m,-\mathbf{k}_j)$. In practice, it means that eigenvectors inconsistent with it should be discarded.

For the odd-parity sector, the eigenvector coefficients $\xi_i(T_0,n_1,\mathbf{k}_1)$, $\xi_i(T_+,n_1,\mathbf{k}_1)$ and $\xi_i(T_-,n_1,\mathbf{k}_1)$ mutually transform under the SU(2) rotation like a spin-1 object. Our recipe is to find a rotation such that $\xi_i(T_+,n_1,\mathbf{k}_1)$ and $\xi_i(T_-,n_1,\mathbf{k}_1)$ vanish. Denote the eigenvector after the rotation as $|\tau_i\rangle=[\tilde{\xi}_i(T_0,n_1,k_1),0,0,\tilde{\xi}_i(T_0,n_2,k_2),0,0,...]^T$. The constraint to $|\tau_i\rangle$ is $\tilde\xi_i(n_m,\mathbf{k}_j)=-e^{i\Theta_{n_m,\mathbf{k}_j}}\tilde\xi_i(n_m,-\mathbf{k}_j)$, in which $\Theta_{n_m,\mathbf{k}_j}$ is the angle between $\langle n_m,\mathbf{k}_j,+|\vec S |n_m,\mathbf{k}_j,+\rangle$ and $\langle n_m,-\mathbf{k}_j,+|\vec S |n_m,-\mathbf{k}_j,+\rangle$ after the same rotations deciding $\tilde\xi_i(T_0, n_m,\mathbf{k}_j)$ and $\tilde\xi_i(T_0, n_m,-\mathbf{k}_j)$. The $\mathcal{T}$ symmetry enforces $\Theta_{n_m,\mathbf{k}_j}=0$ or $\pi$.
\renewcommand{\arraystretch}{1.5}
\begin{longtable*}[t]{>{\centering\arraybackslash}p{0.12\textwidth}>
{\centering\arraybackslash}p{0.18\textwidth}>{\centering\arraybackslash}p{0.12\textwidth}>{\centering\arraybackslash}p{0.13\textwidth}>{\centering\arraybackslash}p{0.13\textwidth}>{\centering\arraybackslash}p{0.13\textwidth}>{\centering\arraybackslash}p{0.13\textwidth}}

    \multicolumn{6}{@{}l}{Table.I Material-dependent parameters for the DFT and DFPT calculations.} \\ \hline \hline
    \textbf{Materials} & \textbf{Energy cutoff (Ry)} & \textbf{Smearing(Ry)} & \textbf{QE $k$-mesh} & \textbf{QE $q$-mesh} & \textbf{EPW $k$-mesh} & \textbf{EPW $q$-mesh} \\ \hline
    \endfirsthead

    \hline  
    Bi$_2$Se$_3$ (slab) & 60 &0.02 & 30$\times$30$\times$1 & 4$\times$4$\times$1 & 180$\times$180$\times$1 & 180$\times$180$\times$1 \\
    Bi$_2$Se$_3$ & 40 &0.01 & 12$\times$12$\times$12 & 3$\times$3$\times$3 & 20$\times$20$\times$20 & 20$\times$20$\times$20 \\
    SnTe & 60 &0.02 & 12$\times$12$\times$12 & 6$\times$6$\times$6 & 20$\times$20$\times$20 & 20$\times$20$\times$20 \\
    Al & 80 &0.02 & 10$\times$10$\times$10 & 10$\times$10$\times$10 & 30$\times$30$\times$30 & 30$\times$30$\times$30 \\
    Pd & 60 &0.02 & 12$\times$12$\times$12 & 6$\times$6$\times$6 & 20$\times$20$\times$20 & 20$\times$20$\times$20 \\
    Cd & 60 &0.02 & 12$\times$12$\times$12 & 6$\times$6$\times$6 & 20$\times$20$\times$20 & 20$\times$20$\times$20 \\
    Ta & 60 &0.02 & 10$\times$10$\times$10 & 10$\times$10$\times$10 & 20$\times$20$\times$20 & 20$\times$20$\times$20 \\
    Pb & 90 &0.01 & 12$\times$12$\times$12 & 6$\times$6$\times$6 & 20$\times$20$\times$20 & 20$\times$20$\times$20 \\
    Hg & 70 &0.02 & 18$\times$18$\times$18 & 3$\times$3$\times$3 & 18$\times$18$\times$18 & 18$\times$18$\times$18 \\
    \hline
    \hline
    \label{m}
\end{longtable*}

\section{Applications}

Below we apply DDM to several representative materials. The input data to construct $\mathbb{V}$ are obtained from QUANTUM ESPRESSO with generalized gradient approximation
 Perdew-Burke-Ernzerhof exchange correlation functional~\cite{giannozzi2017advanced,giannozzi2009quantum,perdew1998perdew}, optimized norm-conserving Vanderbilt pseudopotential~\cite{PhysRevB.88.085117} and Methfessel-Paxton smearing~\cite{PhysRevB.40.3616} for all materials.  The convergence criteria are $10^{-6}(a.u)$ for lattice and atomic relaxations, and $10^{-12}(a.u)$ for electronic iterations. The EPCs from QUANTUM ESPRESSO are further Wannier interpolated by using the EPW code~\cite{ponce2016epw}. Material-dependent setups are summarized in Table \ref{m}. Our DDM code
 to analyze $\mathbb{V}$ is accessible online~\cite{DDMcode}.

\subsection{Bi$_2$Se$_3$ slab}

\begin{figure}[!]
    \centering
    \includegraphics[width=0.5\textwidth]{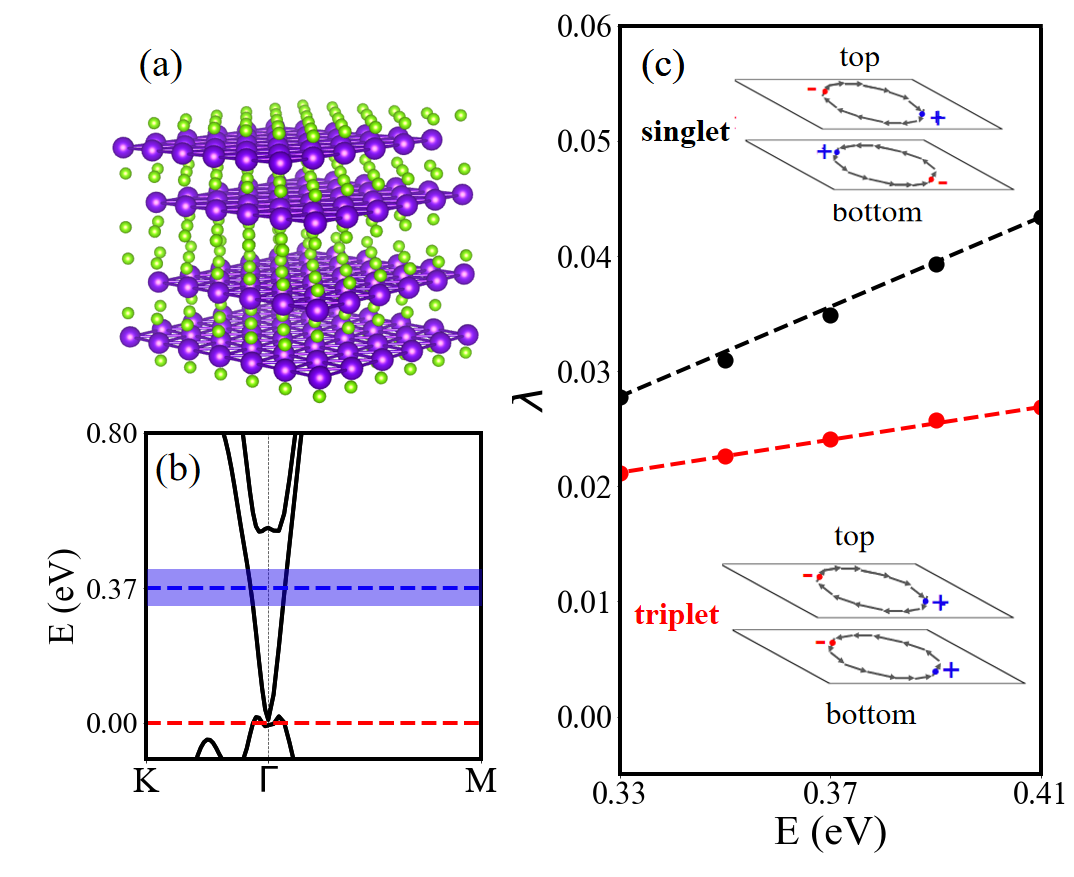}
    \caption{(a) Atomic structure of a two-quintuple layer Bi$_2$Se$_3$ slab employed for the DFT+DFPT calculation; (b) Electron band structure; the shadowed region indicates the range of the electron-doped Fermi level(the bule dash line) for the EPC and DDM calculations. The red dash line means undoped fermi level; (c) Leading pairing strength in the spin-singlet and spin-triplet sectors; the insets schematically decompose the Fermi surfaces into two circles localized on the two surfaces of the slabs. The symmetry of the corresponding DDM eigenvector [cf. Eq. (\ref{eq:eigenvector})] is reflected by plotting the signs of its components at a selected pair of $\pm\mathbf{k}$ points.} 
    \label{fig:my_label}
\end{figure}

We first consider an intuitively transparent example - the well-known 3D topological insulator Bi$_2$Se$_3$~\cite{RevModPhys.83.1057}. By cleaving a slab [Fig.~\ref{fig:my_label}(a)], helical bands form on both surfaces [Fig.~\ref{fig:my_label}(b)]. Within the bulk gap, the helical surface states lead to a doubly-degenerate FS. Since the other bulk bands are fully gapped, the band index $n$ can be omitted. According to our definitions [Eqs. (\ref{eq:T}) and (\ref{eq:newDelta})], $\hat\Delta(\mathbf{k},\pm,\mp)$ refers to intra-surface pairing, and $\hat\Delta(\mathbf{k},\pm,\pm)$ refers to inter-surface pairing.

When the slab is infinitely thick, the two surfaces are decoupled. Therefore, $\langle\hat\Delta(\mathbf{k},\pm,\pm)\rangle=0$, and there is no constraint on the relative phase of $\langle\hat\Delta(\mathbf{k},+,-)\rangle$ and $\langle\hat\Delta(\mathbf{k},-,+)\rangle$, indicating degeneracy between the leading $\langle \hat \Delta^S \rangle$ and $\langle \hat \Delta^{T_0} \rangle$. For a finite thick slab,  Cooper pairs related by spin-conserving inter-surface scatterings, e.g., $\tilde{c}_{\mathbf{-k},+}c_{\mathbf{k},-}\rightarrow \tilde{c}_{\mathbf{k},+}c_{\mathbf{-k},-}$, tend to establish a unified phase to take advantage of the attraction, making $\Delta^S$ the dominating pairing channel. This argument suggests that the spatial separation of the surface states makes the leading pairing channels from the odd-parity and even-parity sectors close in strength. 

Figure ~\ref{fig:my_label}(c) plots the DDM-predicted largest eigenvalues $\lambda_{max}^{S(T)}$ of $\mathbb{V}^{S(T)}$ in a two-quintuple layer Bi$_2$Se$_3$ slab as a function of the Fermi level. In terms of the irreducible representations of the slab point group (D$_{3d}$), the associated eigenstates are assigned to A$_{1g}$ and A$_{1u}$, respectively. The ratio $\lambda_{max}^{T}/\lambda_{max}^{S}$ reaches up to 76.3\%, which pedagogically shows the relevance of EPC calculations for a complete understanding of non-$s$-wave superconductivity. 

We note that pairing requires nonvanishing density of states, so in our calculation the Fermi level is purposely elevated away from the Dirac point [Fig.~\ref{fig:my_label}(b)]. The range of Fermi level for EPC calculations as shadowed in Fig.~\ref{fig:my_label}(b) corresponds to doping 0.043$\sim$0.052 electrons per unit cell, which however is still not enough to support significantly large intrinsic $\lambda$'s. Experimentally, proximity to a bulk superconductor is typically required to achieve surface SC~\cite{wang2012coexistence}.

\begin{figure}[!]
    \centering
    \includegraphics[width=0.5\textwidth]{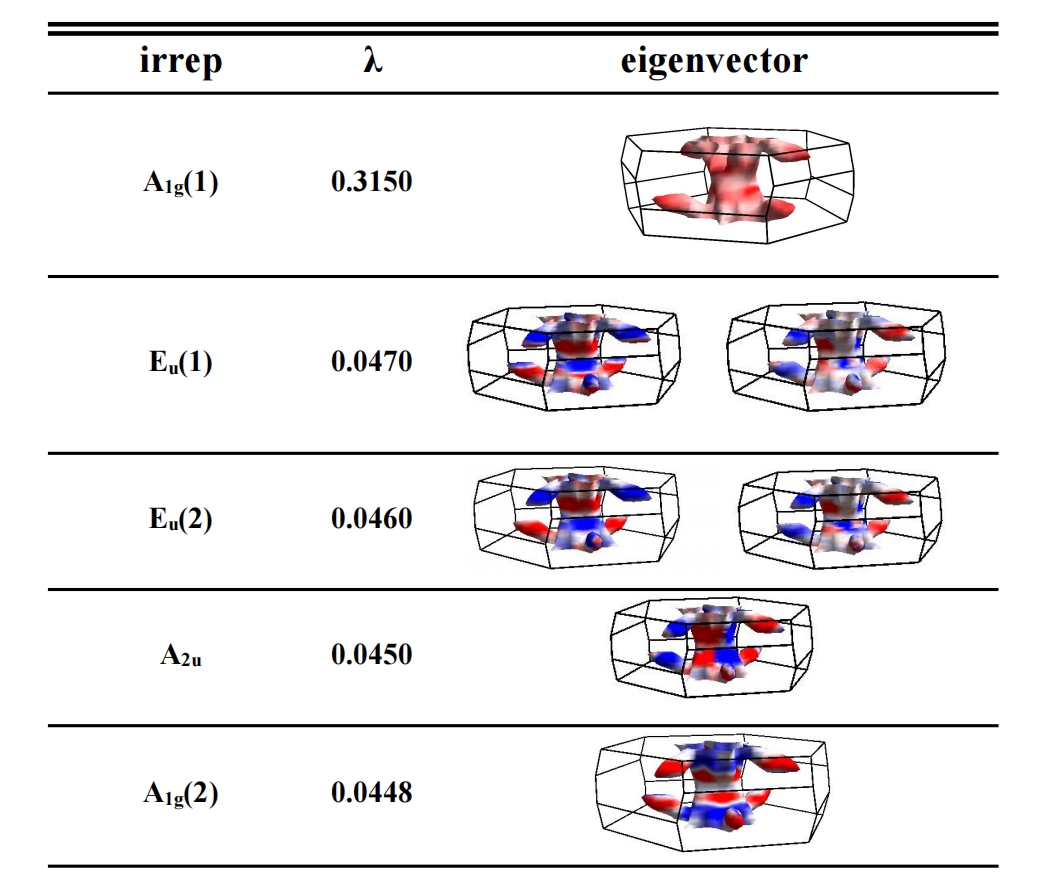}
    \caption{DDM analysis results of electron-doped Bi$_2$Se$_3$. The eigenvalues ($\lambda_i$) and eigenvectors are classified by the irreps. of the D$_{3d}$ point group.  The color contour shows the sign and magnitude of the eigenvector components on the Fermi surface. }
    \label{Bi2Se3}
\end{figure}

\subsection{Doped Bi$_2$Se$_3$ and SnTe}

\begin{figure}[!]
    \centering
    \includegraphics[width=0.5\textwidth]{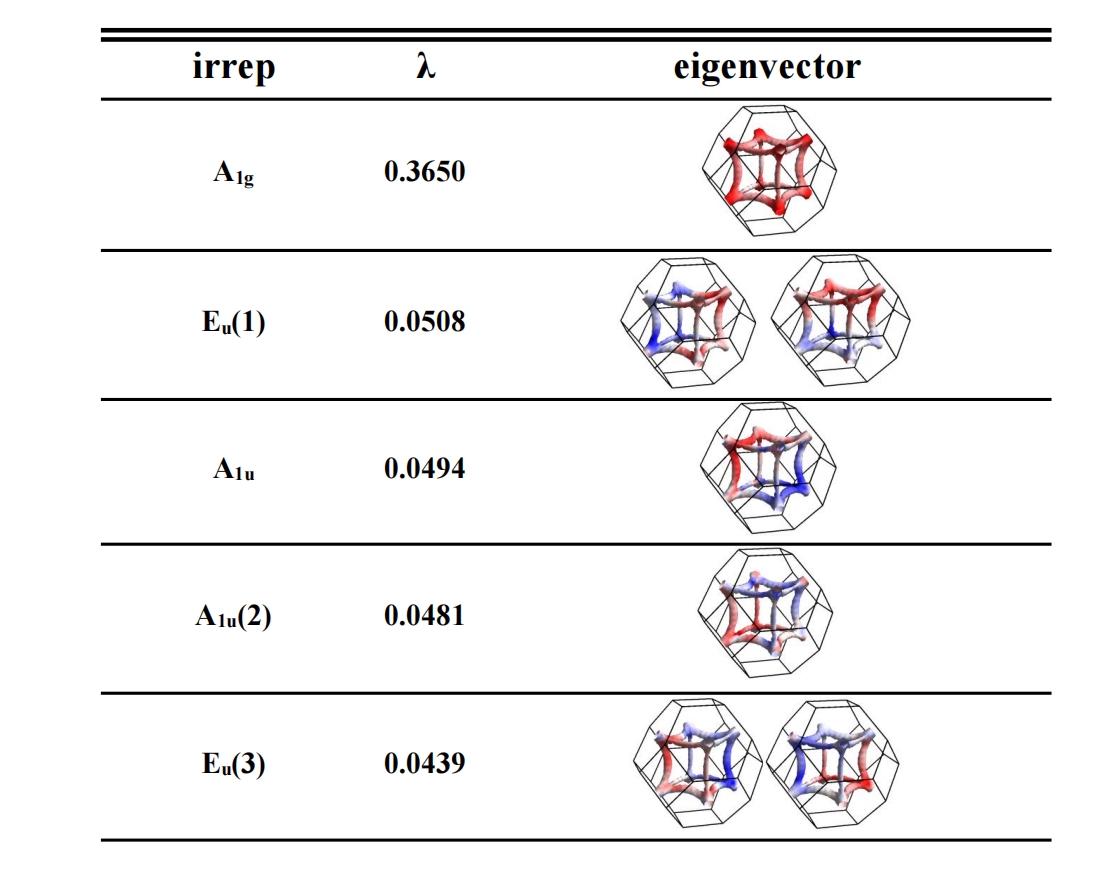}
    \caption{DDM analysis results of electron-doped SnTe. The eigenvalues ($\lambda_i$) and eigenvectors are classified by the irreps. of the O$_h$ point group.  The color contour shows the sign and magnitude of the eigenvector components on the Fermi surface.} 
    \label{fig:SnTe-gap-3}
\end{figure}

The possible topological SC in Cu$_x$Bi$_2$Se$_3$~\cite{hor2010superconductivity,PhysRevLett.107.217001} and Sn$_{1-x}$In$_{x}$Te~\cite{sasaki2012odd}  has motivated several previous EPC calculations on bulk Bi$_2$Se$_3$ and SnTe with charge doping~\cite{zhang2015electron,Wan,Tokura}. Using DDM, we can obtain a systematic differentiation of pairing channels (Figs. \ref{Bi2Se3} and \ref{fig:SnTe-gap-3}). We choose to add 0.12 (0.1) electrons per unit cell to Bi$_2$Se$_3$ (SnTe) in line with the previous calculations to make quantitative comparisons.

For doped Bi$_2$Se$_3$, the leading even-parity pairing $\lambda_{A_{1g}}$ is calculated to be 0.315, in good agreement with the isotropic $\lambda$ reported in Ref.~\cite{zhang2015electron}. The leading odd-parity pairing $\lambda_{Eu}$ is one order of magnitude smaller than $\lambda_{A_{1g}}$, in agreement with Nomoto $et$ $al.$'s qualitative conclusion based on the weak-SOC approximation~\cite{Tokura}. We should mention that Wan and Savrasov~\cite{Wan} in contrast predicted large odd-parity pairings $\lambda_{A_{2u}}$ and $\lambda_{E_{u}}$ based on the FS harmonics projection. This prediction was rationalized by a singular EPC associated with the acoustic phonon at $q$=(0,0,0.04) (in terms of the reciprocal vectors). However, this singular EPC was not reproduced in either Ref.~\cite{zhang2015electron} or \cite{Tokura}. We note that our phonon linewidth calculation gives very similar results to those reported in \cite{zhang2015electron} and \cite{Tokura}. In particular, no singularity is observed at $q$=(0,0,0.04). 


With respect to doped SnTe, DDM predicts that there is no odd-parity pairing channel of the same order of magnitude in strength as the leading even-parity pairing channel. The ratio between $\lambda_{A_{1u}/E_u}$ and $\lambda_{A_{1g}}$ is roughly around 0.1, in agreement with Ref.~\cite{Tokura}. The leading odd-parity gap function shown in Ref.~\cite{Tokura} can also be nicely compared with our Eu(1) eigenstate as plotted in Fig. \ref{fig:SnTe-gap-3}. The other odd-pairing eigenstates of comparative strength have not been shown before. 

\subsection{Elemental metals}
\begin{figure}[!]
    \centering
    \includegraphics[width=0.5\textwidth]{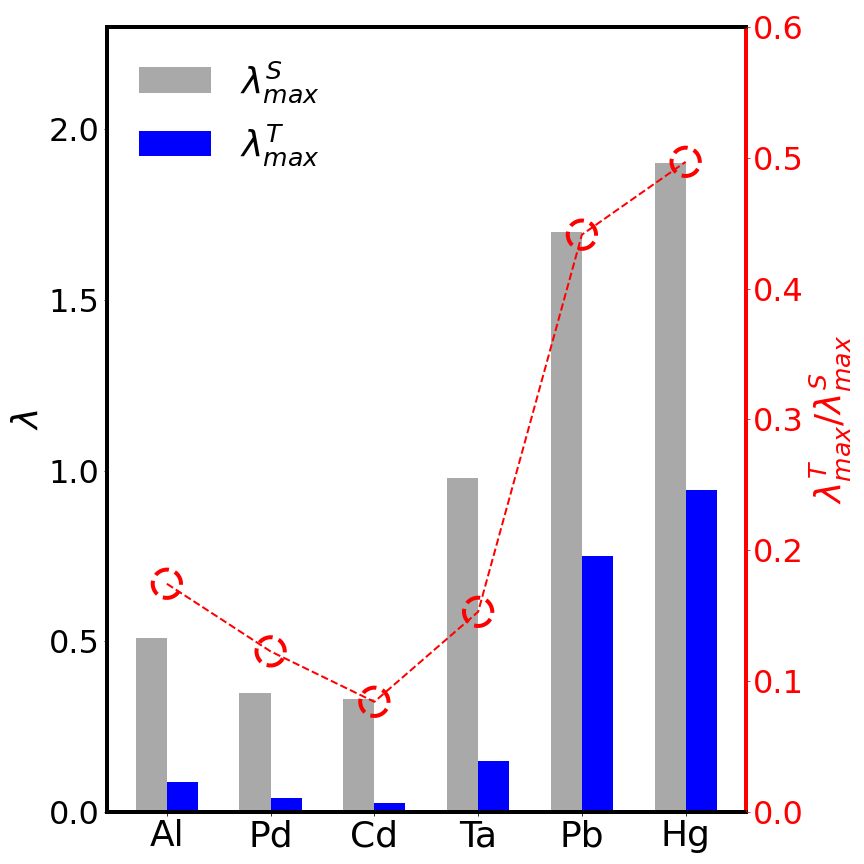}
    \caption{$\lambda_{max}^S$ (grey bar), $\lambda_{max}^T$ (blue bar), and their ratios (red circles) of elemental metals.}
    \label{simple-metal}
\end{figure}

\begin{figure}[h]
    \centering
    \includegraphics[width=0.5\textwidth]{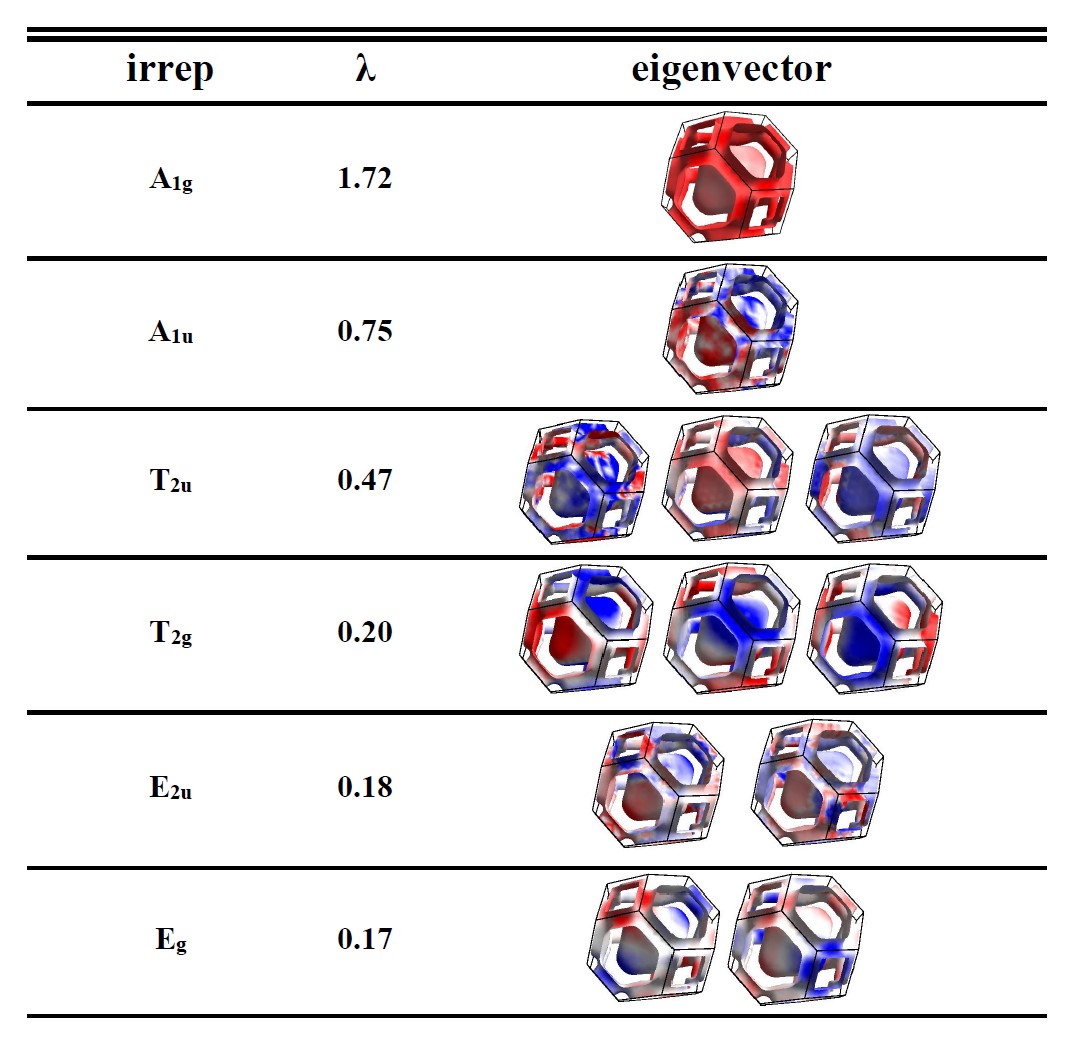}
    \caption{DDM analysis results of Pb. The eigenvalues ($\lambda_i$) and eigenvectors are classified by the irreps. of the O$_h$ point group.  The color contour shows the sign and magnitude of the eigenvector components on the Fermi surface.}
    \label{fig:Pb-gap}
\end{figure}
\begin{figure}[!]
    \centering
    \includegraphics[width=0.5\textwidth]{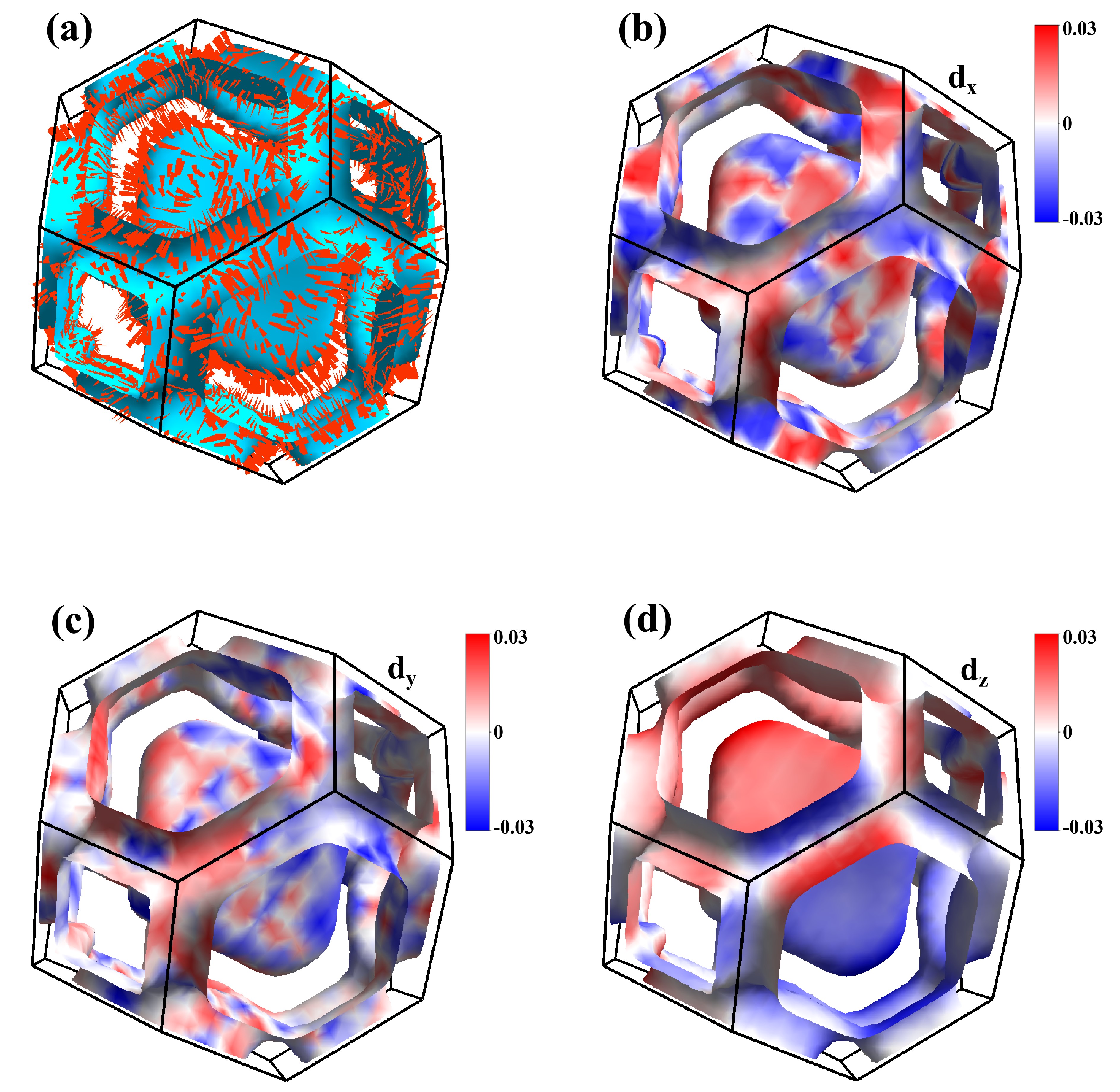}
    \caption{$A_{1u}$ eigenvector of Pb in the d-vector representation: (a) d-vectors on the FS are plotted as arrows; (b)-(d) plot the three components of d-vector, respectively. }
    \label{fig:dv}
\end{figure}
Figure \ref{simple-metal} summarizes $\lambda^{S(T)}_{max}$ of superconducting elemental metals, including Pb,Ta,Al,Pd,Hg,Cd. For all these materials, $\lambda^{S}_{max}$ obtained from the DDM can be well compared with the isotropic dimensionless $\lambda$ reported in literature \cite{PhysRevB.106.L180501,PhysRevB.54.16487,PhysRev.187.525}. The ratio $\lambda_{max}^{T}/\lambda_{max}^{S}$ varies between 0.1 and 0.5. It is interesting to note that the highest ratio occurs in Pb and Hg, in which EPC is known to be quite large. In experiment, Pb is a frequently used component to synthesize potential topological superconductor candidates~\cite{xie2023tuning,pletikosic2023possible,guan2016superconducting,cren2017two}.


Introducing the effect of Coulomb repulsion in general suppresses spin-singlet pairing, making spin-triplet pairing more favorable. We have tested subtracting a constant $\mu^*$(renormalized by a density of states factor) from Eq. (8). We approximate the Coulomb repulsion as momentum independent but pseudospin conserving, i.e., modifying pairing matrix elements with $s_1=s_3$ and $s_2=s_4$ only. Such a real-space delta repulsion effectively suppresses $\lambda_{max}^{S}$ while $\lambda_{max}^T$ is nearly unperturbed. Within this treatment, the critical $\mu^*$ to make $\lambda_{max}^T$ dominate is 0.7 for Pb, and 0.82 for Hg, very close to $\lambda_{max}^{S}$-$\lambda_{max}^T$.

The realistic critical $\mu^*$ could be much lower, because we know that in McMillan’s empirical formula for the conventional singlet pairing, $T_c^S \propto \exp \left[ - \frac{1.04 (1 + \lambda)}{\lambda - \mu^* (1 + 0.62 \lambda)} \right]$ \cite{mcmillan1968transition}. The effect of $\mu^*$ is enhanced by a factor of $(1+0.62\lambda)$. We do not have an empirical formula for the spin-triplet $T_c$ ($T_c^T$). If we naively substitute $\lambda_{max}^{S}$ and $\lambda_{max}^T$ into the same formula and consider $\mu^{S*}=\mu^*$ (effective in leading singlet channel), $\mu^{T*}=0$\cite{PhysRevB.77.094502,Wan} (ineffectice in triplet channel). The critical $\mu^*$ making $T_c^T>T_c^S$ reduces to 0.27 for Pb, and 0.22 for Hg. These values appear more accessible, but we should warn that they are still very crude estimations.

The DDM eigenvectors provide significant physical insights into the associating superconducting states, particularly for understanding the response to an external magnetic field. In Fig. \ref{fig:Pb-gap}, we plot the leading pairing channel of Pb. It is possible to transform these eigenvectors into the d-vector representation commonly used in theoretical literatures \cite{pWaveSuperconductivity}. To do so, we apply an additional SU(2) transformation to rotate the spin expectation value of the Kohn-Sham wavefunctions to align a fixed axis. The same SU(2) transformation rotates the phase-fixed $\xi(T_0,n,\mathbf{k})\sigma_z$ into a non-diagonal 2x2 matrix:

\begin{equation}
\hat\xi’_i(n,\mathbf{k}) =
\begin{pmatrix}
    \xi’_{i,\uparrow',\downarrow'}(n,\mathbf{k}) & \xi'_{i,\uparrow',\uparrow'}(n,\mathbf{k}) \\
    \xi'_{i,\downarrow',\downarrow'}(n,\mathbf{k}) & \xi'_{i,\downarrow',\uparrow'}(n,\mathbf{k})
\end{pmatrix},
\end{equation}
in which the pseudo spin labels $\uparrow',\downarrow'$, have been globally aligned.
The d-vector is then defined as 

\begin{equation}
\mathbf{d}(n,\mathbf{k})=
\begin{pmatrix}
\frac{\xi'_{i,\uparrow',\uparrow'}(n,\mathbf{k})+\xi'_{i,\downarrow',\downarrow'}(n,\mathbf{k})}{\sqrt{2}} \\
\frac{i\xi'_{i,\uparrow',\uparrow'}(n,\mathbf{k})-i\xi'_{i,\downarrow',\downarrow'}(n,\mathbf{k})}{\sqrt{2}} \\
\frac{\xi'_{i,\uparrow',\downarrow'}(n,\mathbf{k})-\xi'_{i,\downarrow',\uparrow'}(n,\mathbf{k})}{\sqrt{2}}
\end{pmatrix}.
\end{equation}

Using the leading spin-triplet channel ($A_{1u}$) of Pb as an example, we make the vector plot of $\mathbf{d}(n,\mathbf{k})$, as well as the color plots of each component in Fig. \ref{fig:dv}.

\section{Conclusion}

In summary, we present an efficient and simple-to-use algorithm for first-principles EPC pairing-channel analysis. It is theoretically known that the leading pairing channel by EPC is $s$-wave\cite{brydon2014odd}. All our first-principles results support this perception. Nevertheless, future material screening in a high-throughput manner might identify candidates in which the $\lambda_{max}^T/\lambda_{max}^S$ is sufficiently high, and there is a chance for the triplet pairing dominates when the electron-electron Coulomb repulsion is also taken into account. These quantifications and further applications in a high-throughput manner are expected to provide useful guide to realize exotic superconductivity.

It is known that by adopting Migdal-Eliashberg theory modern first-principles superconductivity calculations are able to take into account the full frequency dependence of phonon-mediated electron-electron interactions, while the present DDM is formulated at the static mean-field limit. In prospect, the DDM eigen-channels could be used as the initial guess for the self-consistent Migdal-Eliashberg calculation to target unconventional solutions, but a complete formulation requires future efforts.
\section{Acknowledgement}
This work is supported by National Natural Science Foundation of China (Grants No. 92365201, No. 12374062, and
 No. 12474243) and National Key R\&D Program of China 
(2023YFA1406400).

\end{document}